\documentclass[12pt]{article} 

\usepackage[colorlinks=true]{hyperref}
\usepackage[margin=1.in]{geometry}
\usepackage{graphicx}
\usepackage{amsmath}
\usepackage{amssymb}
\usepackage{booktabs}
\usepackage{caption}
\usepackage{subcaption}
\usepackage{verbatim}
\usepackage{float}
\usepackage[english]{babel}
\usepackage{braket}
\usepackage{xy}

\title{Quantum Computing Methods for Supervised Learning}

\author{Viraj Kulkarni$^{1}$, Milind Kulkarni$^1$, Aniruddha Pant$^2$ \\ \\
$^1$ Vishwakarma University\\
$^2$ DeepTek Inc}

\date{\today}

\begin{document}
\hyphenpenalty=1000

\maketitle

\begin{abstract}
\noindent The last two decades have seen an explosive growth in the theory and practice of both quantum computing and machine learning. Modern machine learning systems process huge volumes of data and demand massive computational power. As silicon semiconductor miniaturization approaches its physics limits, quantum computing is increasingly being considered to cater to these computational needs in the future. Small-scale quantum computers and quantum annealers have been built and are already being sold commercially. Quantum computers can benefit machine learning research and application across all science and engineering domains. However, owing to its roots in quantum mechanics, research in this field has so far been confined within the purview of the physics community, and most work is not easily accessible to researchers from other disciplines. In this paper, we provide a background and summarize key results of quantum computing before exploring its application to supervised machine learning problems. By eschewing results from physics that have little bearing on quantum computation, we hope to make this introduction accessible to data scientists, machine learning practitioners, and researchers from across disciplines.
\end{abstract}

\section{Introduction}
Supervised learning is the most commonly applied form of machine learning. It works in two stages. During the training stage, the algorithm extracts patterns from the training dataset that contains pairs of samples and labels and converts these patterns into a mathematical representation called a model. During the inference stage, this model is used to make predictions about unseen samples. Machine learning algorithms, in general, are data hungry; their performance depends heavily on the size of the datasets used for training. However, this training process is computationally expensive, and working with large datasets requires huge amounts of computational horsepower. As the number and volume of datasets available for research and commercial purposes continues to grow exponentially, new technologies such as NVIDIA CUDA \cite{nickolls2008scalable} and Google TPUs \cite{jouppi2017datacenter} have emerged to enable faster processing of data. Computational speeds have been increasing rapidly over the last several decades in accordance with the observation that the number of transistors in a dense integrated circuit (IC) doubles about every two years - this is informally known as Moore’s Law \cite{friedman2015moore}. The semiconductor manufacturing process has shrunk from 10 micrometers in 1970 to about 5 nanometers in 2020. It is believed, however, that we are nearing the limits of the processing power that classical computers can offer us \cite{leiserson2020there}. At scales smaller than this, quantum mechanical effects come into play \cite{sperling2018quantum}, and they impose physical limitations on how small electronic components can get.

Quantum computing, on the other hand, proposes to leverage these quantum mechanical effects to carry out computation. In contrast to classical computers that operate on bits that can exist in only one out of two states at a time, quantum computers exploit the fact that quantum bits (qubits) can exist in any one of the infinite possible linear superpositions of these two states. This allows quantum computers to execute multiple paths of computation simultaneously. Quantum computers can efficiently solve computational problems which are believed to be intractable for classical computers \cite{nielsen2002quantum}. Owing to its roots in theoretical physics, most research articles on the topic are written for physicists; this makes them difficult to access for researchers from other fields. The purpose of this paper is to give a broad overview of the synergies between quantum computing and machine learning. We briefly outline the history of quantum physics, describe the preliminaries of quantum computation, and then review the latest research on applying the principles of quantum computation to supervised machine learning. A large-scale general-purpose quantum computer does not yet exist, but restricted quantum machines capable of solving optimization problems are already being sold commercially \cite{castelvecchi2017ibm}\cite{johnston2013d}. By eschewing results from physics that have little bearing on quantum computation and by providing additional background that may benefit the unfamiliar reader, we hope to make this introduction accessible to data scientists, machine learning practitioners, and researchers from other fields. 

Quantum mechanics arose through a number of discoveries in the early twentieth century. Einstein \cite{einstein17heuristic} explained the photoelectric effect in 1905 by postulating that light and all electromagnetic radiation is made up of discrete particles that later came to be called photons. Previously in 1865, Maxwell \cite{maxwell1865viii} had demonstrated that electric and magnetic fields travel through space as waves. De Broglie \cite{broglie1924xxxv} proposed in 1923 that particles can exhibit wave-like properties and waves can behave like particles. Building on his approach, Heisenberg \cite{edwards1979mathematical} developed matrix mechanics, and Schrödinger \cite{schrodinger1926undulatory} developed wave mechanics, both of which were later found to be equivalent. These developments laid the foundation of quantum mechanics. The equations of quantum mechanics have since been extensively tested in innumerable experiments, but even after almost a century of debate, physicists strongly disagree over how those equations should be interpreted and mapped to reality \cite{schlosshauer2013snapshot}.

Benioff \cite{benioff1980computer} in 1980 and Feynman \cite{feynman1999simulating} in 1982 observed that simulating the evolution of certain quantum systems may be an intractable problem that cannot be solved efficiently by computers, and yet, these quantum systems solved the problem by merely evolving thus suggesting that the evolution of quantum systems could be used as a method of computation. In 1985, Deutsch \cite{deutsch1985quantum} designed a universal quantum computer as a quantum-counterpart to the Universal Turing Machine. Deutsch and Jozsa \cite{deutsch1992rapid} proposed the Deutsch-Jozsa problem for which the deterministic quantum algorithm is exponentially faster than any deterministic classical solution. Shor's algorithm \cite{shor1999polynomial} for factoring large integers was the first quantum algorithm that could solve a problem of practical significance faster than any classical algorithm. Grover's algorithm \cite{grover1996fast} showed quadratic improvement in unordered search problems. These results laid the foundation of quantum computation. Since then, quantum algorithms have been proposed for numerous areas including cryptography, search and optimisation, simulation of quantum systems, solving large systems of linear equations, and machine learning \cite{montanaro2016quantum}.

The benefits quantum computing can bring to machine learning go beyond speed-up in execution. Many tasks in machine learning such as maximum likelihood estimation using hidden variables, principal component analysis, training of neural networks etc. require optimization of a non-convex objective function. Optimizing non-convex functions is an NP-hard problem. Classical optimization methods such as gradient descent can get stuck at local minimum or saddle points and may never find the global minimum. Adiabatic quantum computers use quantum annealing to solve non-convex optimization problems by finding low-energy configurations of an appropriate energy function by exploiting tunneling effects to escape local minima \cite{santoro2006optimization}. Methods based on Grover's search can find the global minimum in a discrete unordered search space. Many machine learning algorithms involve repeated execution of linear algebra routines on large matrices. Quantum solutions can offer exponential speed-up for these routines \cite{harrow2009quantum}\cite{wiebe2012quantum}. Besides optimizing subroutines used in classical machine learning algorithms, many fully quantum versions of these algorithms have also been developed.

The term \textit{quantum machine learning} is generally used to denote analysis of classical data on quantum computers. This is known as \textit{quantum-enhanced} machine learning. There are however other ways in which the fields of quantum computing and machine learning overlap. Classical machine learning can be applied to data emanating from quantum systems to solve problems in physics. Another stream of research deals with generalizing classical machine learning to work with quantum data where the input and output are quantum states. Recently, Tang \cite{tang2019quantum} developed a classical algorithm for recommendation systems that was inspired by quantum computing creating a new category referred to as \textit{quantum-inspired} algorithms. These are classical algorithms that can be run on conventional computers which borrow ideas from quantum computing to achieve significant theoretical speed-ups over the best prevailing classical algorithms. This paper limits itself to quantum-enhanced machine learning and presents a selection of quantum approaches for implementing supervised machine learning algorithms on quantum computers. Far from being a comprehensive review of the field, it aims to offer the reader a background on the multitude of approaches proposed over the years with enough detail to set the stage for a more detailed exploration. For additional information, we recommend the following excellent surveys and reviews: \cite{schuld2015introduction}\cite{wittek2014quantum}\cite{biamonte2017quantum}\cite{adcock2015advances}\cite{arunachalam2017guest}\cite{kopczyk2018quantum}\cite{dunjko2018machine}\cite{dunjko2020non}\cite{montanaro2016quantum}.

\section{Background of Quantum Computation}
Quantum mechanics is based on four fundamental postulates \cite{dunjko2018machine}\cite{nielsen2002quantum}: (1) the pure state of a quantum system is given by a unit vector $\ket{\psi}$ in a complex Hilbert space; (2) the evolution of a pure state in a closed system is governed by a Hamiltonian $H$ as specified by Schr\"{o}dinger's equation $H \ket{\psi} = i \hbar \frac{\partial}{\partial t} \ket{\psi}$; (3) the quantum state of a composite system is given by the tensor product of the individual systems; (4) projective measurements (observables) are specified by Hermitian operators, and the process of measurement changes the observed system from $\ket{\psi}$ to an eigenstate $\ket{\phi}$ with probability given by the Born rule $p(\phi) = |\langle{\psi} \ket{\phi} |^2$. In this section, we briefly set up the background of quantum computation based on the above postulates.

\subsection{Single Qubit}
A classical bit can exist in one of two states denoted as 0 and 1. A quantum bit or qubit can exist not only in these two discrete states but in all possible linear superpositions of them. Mathematically, the quantum state of a qubit is represented as a state vector in a two-dimensional Hilbert space. In the Dirac notation, the state vector of a qubit $\psi$ is called a \textit{ket} and is written as:
\begin{align}
\ket{\psi} = \alpha \ket{0} + \beta \ket{1}
\end{align}
where $\alpha$ and $\beta$ are complex numbers and $|\alpha|^2+|\beta|^2=1$. The Born's rule tells us that if this qubit is measured, we will get $\ket{0}$ with probability $|\alpha|^2$ and $\ket{1}$ with probability $|\beta|^2$. Quantum measurements are non-deterministic, and the act of measurement changes the quantum state irreversibly. Before measurement, the qubit exists in a quantum superposition of the states $\ket{0}$ and $\ket{1}$. The outcome of the measurement, however, is not quantum but classical i.e. you get either a $\ket{0}$ or a $\ket{1}$ but not a superposition of the two. During the measurement, the quantum state collapses\footnote{Different interpretations exist regarding the collapse of the quantum state \cite{schlosshauer2013snapshot}. The popular Copenhagen Interpretation suggests that the wave function of a quantum system collapses on observation. The alternative Many Worlds Interpretation suggests that there is no collapse of the wave function; instead, the act of observation results in the observer getting entangled with the observed system.} to the classical state it gets observed in, and all subsequent measurements deterministically result in this same outcome with a probability equal to 1.

The choice of basis vectors $\ket{0}$ and $\ket{1}$ is arbitrary. We can represent the system using a different set of orthogonal basis vectors such as $\ket{+}$ and $\ket{-}$ (called the Hadamard or sign basis). Once the computational basis is decided, kets can be represented as column vectors:
\begin{align}
\ket{0} = \begin{bmatrix} 1 \\ 0 \end{bmatrix}, 
\ket{1} = \begin{bmatrix} 0 \\ 1 \end{bmatrix}
\end{align}
\begin{align}
\ket{\psi} = \alpha \begin{bmatrix} 1 \\ 0 \end{bmatrix} + \beta \begin{bmatrix} 0 \\ 1 \end{bmatrix} = \begin{bmatrix} \alpha \\ \beta \end{bmatrix}
\end{align}
The two representations for $\ket{\psi}$ given in equations (1) and (3) are equivalent.

\subsection{Multiple Qubits}
The quantum state of a system consisting of more than one \textit{unentangled} qubits can be represented as the tensor product of the quantum states of the individual qubits. The state of a two-qubit system comprising of qubits represented by $\ket{\psi_1}$ and $\ket{\psi_2}$ can be written as $\ket{\Psi} = \ket{\psi_1} \otimes \ket{\psi_2}$. In general, the state of $n$ qubits $\{\ket{\psi_1}, \ket{\psi_2}, \dots, \ket{\psi_n}\}$ can be represented as:
\begin{align}
\ket{\Psi}=\ket{\psi_1} \otimes \ket{\psi_2} \otimes \dots \otimes \ket{\psi_k}=\ket{\psi_1\psi_2\dots\psi_k} \,.
\end{align}
However, not all multi-qubit states can be represented as a tensor product of individual states. Consider the state below, one of the Bell states:
\begin{align}
\ket{\Phi^+} = \frac{1}{\sqrt{2}}(\ket{00} + \ket{11})
\end{align}
Suppose it could be decomposed into the tensor product of two states as below:
\begin{align}
\ket{\Phi^+} &= (a_1\ket{0} + b_1\ket{1}) \otimes (a_2\ket{0} + b_2\ket{1}) \nonumber \\
&= a_1a_2\ket{00} + a_1b_2\ket{01} + a_2b_1\ket{10} + b_1b_2\ket{11}
\end{align}
From equations (5) and (6), we know that $a_1b_2 = a_2b_1 = 0$. Therefore, either $a_1a_2 = 0$ or $b_1b_2 = 0$. But, from equation (5), both $a_1a_2 \ne 0$ and $b_1b_2 \ne 0$. This proves that the Bell state $\ket{\Phi^+}$ cannot be decomposed into the tensor product of two single-qubit states. In such cases, we say that the two qubits are \textit{entangled}. Given an entangled pair of qubits, measurement on one qubit instantaneously affects the other qubit. Entanglement plays a central role in many quantum algorithms especially in the field of quantum cryptography. There is no counterpart to quantum entanglement in classical physics.

\subsection{Quantum Gates}
Classical computers manipulate information stored in bits using logic gates such as AND, OR, NOT, NAND, XOR etc. Likewise, quantum computers manipulate qubits using quantum gates. Transformations on quantum states are represented as rotation of the Hilbert space. Rotation is linear and reversible. Consequently, all transformations on quantum states must be linear and reversible. Quantum gates essentially transform the system from one state to another state. These transformations can be represented as matrices. The simplest quantum gate is the NOT gate. The NOT gate transforms $\ket{\psi_1} = \alpha \ket{0} + \beta \ket{1}$ to $\ket{\psi_2} = \alpha \ket{1} + \beta \ket{0}$ and can be represented as:
\begin{align}
\textit{NOT}=\begin{bmatrix} 0 & 1 \\ 1 & 0 \end{bmatrix} \ \label{eq:not}
\end{align}

The Hadamard gate acts on a single qubit. It is often used to map a qubit from one of its basis states into an equal superposition of all basis states. It transforms $\ket{0}$ to \mbox{$\frac{1}{\sqrt{2}}(\ket{0}+\ket{1})$} and $\ket{1}$ to \mbox{$\frac{1}{\sqrt{2}}(\ket{0}-\ket{1})$} and is given by:
\begin{align}
H = \frac{1}{\sqrt{2}} \begin{bmatrix} 1 & 1 \\ 1 & -1 \end{bmatrix}
\end{align}
In general, an n-qubit Hadamard gate is used to initialize an n-qubit system into an equal superposition of all basis states:
\begin{align}
\ket{0^n} \xrightarrow{\text{$H^{\otimes n}$}} \frac{1}{2^{n/2}}\sum_{x\in\{0,1\}^n}\ket{x}
\end{align}
where $x\in\{0,1\}^n$ denotes all strings of length $n$ consisting of $0$ and $1$.

The CNOT (controlled-NOT) gate acts on two qubits where the first qubit acts as a control signal that decides whether the NOT operation should be performed on the second qubit. If the control qubit is $\ket{1}$, the NOT operation is applied; if it is $\ket{0}$, it is not applied. The CNOT gate leaves the states ${\ket{00}}$ and ${\ket{01}}$ unchanged, while it maps ${\ket{10}}$ to ${\ket{11}}$ and ${\ket{11}}$ to ${\ket{10}}$. It is represented as:
\begin{align}
CNOT = \begin{bmatrix} 1 & 0 & 0 & 0 \\ 0 & 1 & 0 & 0 \\ 0 & 0 & 0 & 1 \\ 0 & 0 & 1 & 0 \end{bmatrix}
\end{align}

The SWAP gate swaps the states of two qubits transforming $\ket{\psi,\phi}$ to $\ket{\phi,\psi}$. The CSWAP (controlled-SWAP) gate acts on three qubits and swaps the state of the second and third qubit if the first qubit is $\ket{1}$. The Toffoli gate (CCNOT) acts on three qubits and performs the computation:
\begin{align}
\ket{a, b, c} \rightarrow \ket{a, b, c \oplus ab}
\end{align}

\subsection{Quantum Parallelism}
While classical computers can execute only one computational path at a time, quantum computers can leverage the ability of quantum states to exist in superpositions to simultaneously execute multiple computational paths. For example, consider the classical function $f(x): \{0,1\}^2 \rightarrow \{0,1\}$. The function takes two bits as input and outputs a single bit. To evaluate $f$ on all two-bit permutations using classical computation, we need to call $f$ four times: $f(0,0)$, $f(0,1)$, $f(1,0)$, and $f(1,1)$. Quantum superposition allows us to evaluate all four inputs in a single call to $f$.

Since quantum transformations must be reversible and $f$ is not reversible, we define a reversible quantum function:
\begin{align}
U_f\ket{x}\ket{y}\rightarrow\ket{x}\ket{y \oplus f(x)}
\end{align}
The input $\ket{\phi}$ is set up in a superposition of states by initializing two qubits to $\ket{0}$ and applying the Hadamard transform:
\begin{align}
\ket{\phi} = (H \otimes H)\ket{00} = \frac{\ket{0}+\ket{1}}{\sqrt{2}} \otimes \frac{\ket{0}+\ket{1}}{\sqrt{2}} = \frac{1}{2}(\ket{00}+\ket{01}+\ket{10}+\ket{11})
\end{align}
Setting $\ket{y} = \ket{0}$, we apply $U_f$ as follows:
\begin{align}
U_f\ket{\phi}\ket{0}&=\frac{1}{2}U_f(\ket{00}+\ket{01}+\ket{10}+\ket{11})\otimes\ket{0} \nonumber \\
&= \frac{1}{2}U_f(\ket{00,0}+\ket{01,0}+\ket{10,0}+\ket{11,0}) \nonumber \\
&= \frac{1}{2}(\ket{00,f(00)}+\ket{01,f(01)}+\ket{10,f(10)}+\ket{11,f(11)})
\end{align}
Thus, with a single application of $f$, we simultaneously evaluate four inputs. Using the Hadamard transform the set up in the input in an equal superposition of all basis vectors is a useful starting point for many quantum algorithms.

\subsection{No Cloning Theorem}
An important result that has profound implications is the \textit{no cloning theorem} \cite{wootters1982single} \mbox{which states} that it is not possible to create a copy of an unknown quantum state. Since measurement irreversibly changes the quantum state, given a single copy of the state \mbox{$\ket{\psi} = \alpha \ket{0} + \beta \ket{1}$}, the values of the amplitudes $\alpha$ and $\beta$ cannot be exactly determined. Although quantum parallelism can be leveraged to simultaneously execute multiple computational paths, the no cloning theorem places restrictions on the amount of information one can extract from the final quantum state\footnote{Although perfect cloning is impossible, Buzek and Hillery \cite{buvzek1996quantum} proposed a universal cloning machine that can make imperfect copies of unknown quantum states with high fidelity.}.

\subsection{Adiabatic Quantum Computation}
Numerous models have been proposed for quantum computing such as the quantum Turing machine, quantum circuit model, adiabatic quantum computing, measurement-based quantum computing, blind quantum computing, topological quantum computing etc. \cite{dunjko2018machine}. All these models are computationally equivalent, but they are implemented very differently. An approach that has shown promise in solving optimization problems is adiabatic\footnote{In the context of quantum computing, an adiabatic process is a process which changes the state of a system so gradually that the state can adapt its configuration at each point.} quantum computing \cite{farhi2000quantum} and is of particular interest because building restricted quantum computers to perform \textit{quantum annealing} (section 3.2) based on adiabatic quantum computing is simpler than building universal quantum computers. In adiabatic quantum computing, the optimization problem to be solved is encoded as a boolean satisfiability problem such that the ground state of its Hamiltonian\footnote{The Hamiltonian operator represents the total energy of the system.} represents the desired solution. The quantum system is initially set up with a simple Hamiltonian that is easy to construct. The system is then evolved from the initial state to a final state. The adiabatic theorem states that if the system is evolved slowly enough, it will remain in the ground state of the instantaneous Hamiltonian throughout the evolution. The final system configuration then represents the solution to the optimization problem.

\section{Quantum Machine Learning}
Quantum computing methods for machine learning can be divided into two broad classes: \mbox{(1) methods} designed to run on a universal quantum computer that involve preparation, storage, and processing of quantum states and the retrieval of the classical solution from these states; (2) methods designed to run on \textit{quantum annealers} that solve optimization problems through the physical evolution of quantum systems according to the principles of adiabatic quantum computing. In section 3.1, we describe common subroutines designed for circuit quantum computers that can be applied to machine learning problems. In section 3.2, we present quantum annealing that can solve quadratic unconstrained binary optimization (QUBO) problems. Finally, in section 3.3, we present quantum versions of selected classical machine learning algorithms.

\subsection{Important Subroutines of Quantum Algorithms}
A straightforward approach to achieving speed-ups over classical machine learning algorithms is to identify their computationally expensive and frequently executed subroutines and develop quantum alternatives for them. In this section, we describe some common subroutines that form a part of many quantum learning algorithms.

\subsubsection{Quantum Encoding}
Qubits are a scarce and expensive resource. The restricted physical implementations of quantum computers available today have very few qubits\footnote{The recent \textit{quantum supremacy} experiment conducted by Google used only 54 qubits \cite{arute2019quantum}.}. An important question therefore that has implications for performance and feasibility is how to represent classical data in quantum states. Suppose we have a dataset $D$ of $N$ instances:
\begin{align}
D=\{x_1,...,x_i,...,x_N\}
\end{align}
where each $x_i$ is a real number. In \textit{basis encoding}, each instance $x_i$ is encoded as $\ket{b_i}$ where $b_i$ is the binary representation of $x_i$. The dataset $D$ can then be represented as a superposition of all computational basis states:
\begin{align}
\ket{\psi} = \frac{1}{\sqrt{N}}\sum_{i=1}^{N}\ket{b_i}
\end{align}
In \textit{amplitude encoding}, data is encoded in the amplitudes of quantum states. The above dataset $D$ can be encoded as:
\begin{align}
\ket{\psi} = \frac{\sum_{i=1}^{N}x_i\ket{i}}{||D||}
\end{align}
Besides basis and amplitude encoding, many other methods of encoding exist such as \textit{ Qsample encoding}, \textit{dynamic encoding}, \textit{squeezing embedding}, \textit{displacement embedding}, \textit{Hamiltonian encoding} etc. \cite{schuld2019quantum}\cite{lloyd2020quantum}.

\subsubsection{Grover's Algorithm and Amplitude Amplification}
Grover's algorithm \cite{grover1996fast} is a quantum search algorithm that offers a quadratic speed-up over classical algorithms when performing a search over an unstructured search space. Suppose we are given a set of $N$ elements $X=\{x_1,...,x_i,...,x_N\}$ where $x_i \in \{0,1\}^m$ and a boolean function $f:\{0,1\}^m \rightarrow\{0,1\}$ such that:
\begin{align}
f(x)=
\begin{cases}
1 & x = x^*\\
0 & x \neq x^*
\end{cases} \,
\end{align}
Any classical algorithm that performs a search for $x^*$ in $X$ is $O(N)$ in time. Grover's algorithm can perform such a search in $O(\sqrt{N})$. The algorithm has three steps to it.

In the first step, a quantum state is set up in an equal superposition of basis states using the Hadamard transform. As an example, consider $N=8$. We set up the state using 3 qubits as:
\begin{align}
\ket{\psi_1}&=(H \otimes H \otimes H)\ket{000} \nonumber \\
&=\frac{1}{2\sqrt{2}} (\ket{000}+\ket{001}+\ket{010}+\ket{011}+\ket{100}+\ket{101}+\ket{110}+\ket{111}) \nonumber \\
&=\begin{bmatrix} \frac{1}{2\sqrt{2}} & \frac{1}{2\sqrt{2}} & \frac{1}{2\sqrt{2}} & \frac{1}{2\sqrt{2}} & \frac{1}{2\sqrt{2}} & \frac{1}{2\sqrt{2}} & \frac{1}{2\sqrt{2}} & \frac{1}{2\sqrt{2}}\end{bmatrix}^T
\end{align}

The second step referred to as \textit{phase inversion} deals with flipping the amplitude of each $\ket{x}$ if $f(x)=1$ and leaving it unchanged if $f(x)=0$. To do this, we define a unitary quantum oracle $O\ket{x}=(-1)^{f(x)}\ket{x}$. Suppose, in our example, $x^*$ is present at the fourth position. Applying gate $O$ on $\ket{\psi_1}$ gives us:
\begin{align}
\ket{\psi_2}=\begin{bmatrix} \frac{1}{2\sqrt{2}} & \frac{1}{2\sqrt{2}} & \frac{1}{2\sqrt{2}} & \frac{-1}{2\sqrt{2}} & \frac{1}{2\sqrt{2}} & \frac{1}{2\sqrt{2}} & \frac{1}{2\sqrt{2}} & \frac{1}{2\sqrt{2}}\end{bmatrix}^T
\end{align}

The third step referred to as \textit{inversion around the mean} involves flipping all amplitudes around their collective mean $\mu=\frac{1}{N}\sum_x{\alpha_x}$. This is performed by the \textit{Grover diffusion operator} $G$:
\begin{align}
\sum_x{\alpha_x\ket{x}} \xrightarrow{\text{$G$}} \sum_x{(2\mu-\alpha_x)\ket{x}}
\end{align}
Applying $G$ to $\ket{\psi_2}$ gives us:
\begin{align}
\ket{\psi_3}=\begin{bmatrix} \frac{1}{4\sqrt{2}} & \frac{1}{4\sqrt{2}} & \frac{1}{4\sqrt{2}} & \frac{5}{4\sqrt{2}} & \frac{1}{4\sqrt{2}} & \frac{1}{4\sqrt{2}} & \frac{1}{4\sqrt{2}} & \frac{1}{4\sqrt{2}}\end{bmatrix}^T
\end{align}

Thus, after one iteration, the amplitude of the target element $x^*$ is higher than the amplitudes of other elements. If we were to measure the system at this point, we would get the target element as outcome with a probability of $(\frac{5}{4\sqrt{2}})^2=78\%$. The second and third steps are repeated $\sqrt{N}$ times to maximize this probability. After the second iteration, we get the below state which will find the target element with a probability of $95\%$:
\begin{align}
\ket{\psi_4}=\begin{bmatrix} \frac{-1}{8\sqrt{2}} & \frac{-1}{8\sqrt{2}} & \frac{-1}{8\sqrt{2}} & \frac{11}{8\sqrt{2}} & \frac{-1}{8\sqrt{2}} & \frac{-1}{8\sqrt{2}} & \frac{-1}{8\sqrt{2}} & \frac{-1}{8\sqrt{2}}\end{bmatrix}^T
\end{align}

The same algorithm can also find $k$ matching entries instead of a single entry. Several modifications have been proposed that extend this work. Durr and Hoyer \cite{durr1996quantum} propose a quantum algorithm to find the index of the minimum value from a list of $N$ values in time $O(c\sqrt{N})$ with a probability of at least $(1-\frac{1}{2^c})$. These methods generalize Grover's search and are collectively referred to as \textit{amplitude amplification} techniques \cite{brassard1997exact}.

\subsubsection{Calculating Inner Products using Swap Test}
The swap test \cite{buhrman2001quantum} is a simple subroutine used to compute the overlap between two quantum states $\ket{\phi}$ and $\ket{\psi}$. Quantum procedures can be easily described using circuit diagrams. The circuit diagram of the swap test is shown in figure 1. 
\begin{figure}
\centering
\includegraphics[keepaspectratio,height=3cm, width=6cm]{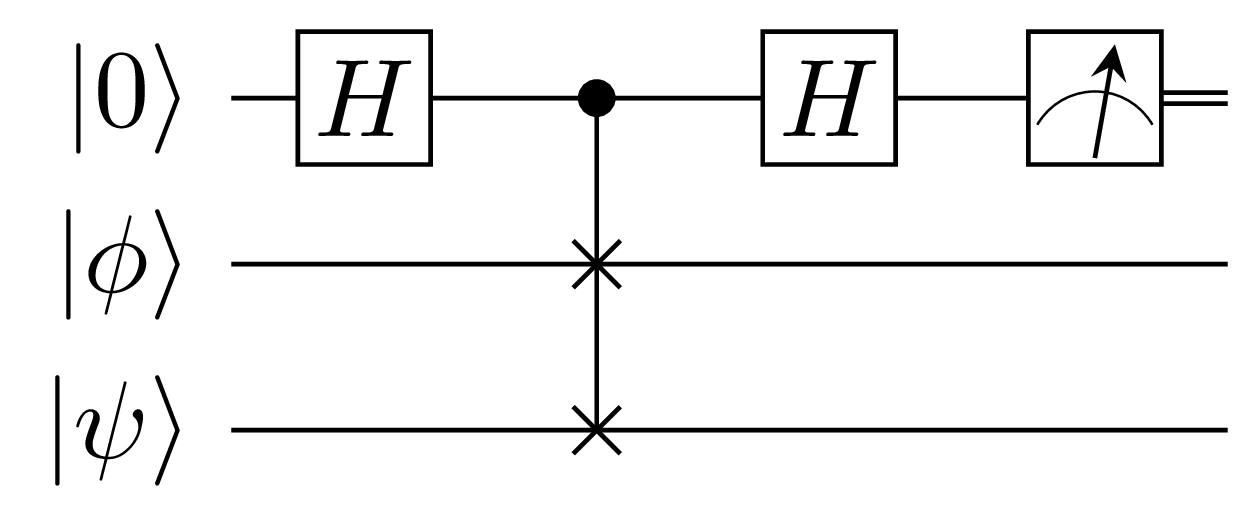}
\caption{Circuit diagram describing swap test. Computation proceeds from left to right.}
\end{figure}
The system is initially prepared in the state $\ket{0,\phi,\psi}$. The Hadamard gate applied on the first \textit{ancilla qubit}\footnote{Input qubits that do not hold any input data but are added to satisfy other conditions (most often reversibility of the transformation) are called auxiliary or ancillary qubits.} transforms the state to $\frac{1}{\sqrt{2}}(\ket{0,\phi,\psi}+\ket{1,\phi,\psi})$. The CSWAP further transforms it to $\frac{1}{\sqrt{2}}(\ket{0,\phi,\psi}+\ket{1,\psi,\phi})$. After the application of the second Hadamard gate to the first qubit, the state can be written as $\frac{1}{2}\ket{0}(\ket{\phi,\psi}+\ket{\psi,\phi})+\frac{1}{2}\ket{1}(\ket{\phi,\psi}-\ket{\psi,\phi})$. The probability of measuring the first qubit as $\ket{0}$ is given by $p=\frac{1}{2}+\frac{1}{2}{|\braket{\psi|\phi}|}^2$. If $\ket{\phi}$ and $\ket{\psi}$ are equal, ${|\braket{\psi|\phi}|}^2=1$, and the observed value of $p$ is 1. If $\ket{\phi}$ and $\ket{\psi}$ are orthogonal, ${|\braket{\psi|\phi}|}^2=0$, and the observed value of $p$ is $\frac{1}{2}$. The degree of overlap given by the inner product of the two states can be estimated with this method to precision $\epsilon$ using $O(1/\epsilon)$ copies \cite{dunjko2018machine}.

\subsubsection{Solving Systems of Linear Equations (HHL)}
Solving a system of linear equations is an important problem ubiquitous throughout science and engineering. A seminal result in quantum computation is the HHL algorithm \cite{harrow2009quantum} that solves the following problem: given a Hermitian matrix $A \in \mathbb{R}^{N \times N}$ and a unit vector $\overrightarrow{b} \in \mathbb{R}^N$, find a solution vector $\overrightarrow{x} \in \mathbb{R}^N$ that satisfies the equation $A\overrightarrow{x}=\overrightarrow{b}$.

We present here a condensed outline of the algorithm. The solution we are interested in is $\ket{x}=A^{-1}\ket{b}$. Let $\{v_1,...,v_N\}$ be the eigenvectors of $A$ with corresponding eigenvalues $\{\lambda_1,...,\lambda_N\}$. The vector $\overrightarrow{b}$ is encoded using amplitude encoding (section 3.1.1) as \mbox{$\ket{b}=\sum_{i=1}^{N}{\beta_i\ket{v_i}}$}. \textit{Hamiltonian simulation} is used to transform the matrix $A$ into a unitary operator, and \textit{quantum phase estimation}\footnote{The quantum phase estimation algorithm can estimate the phase (or eigenvalue) of an eigenvector of a unitary operator.} is used to carry out eigendecomposition to get the state $\sum_{i=1}^{N}{\beta_i\ket{v_i}\ket{\lambda_i}}$. An ancilla qubit is added, rotation conditioned on $\ket{\lambda_i}$ is carried out, and the eigenvalue register $\ket{\lambda_i}$ is uncomputed to yield a state proportional to \mbox{$\sum_{i=1}^{N}{\beta_i \lambda_i^{-1}\ket{v_i}} = A^{-1}\ket{b} = \ket{x}$.}

It is important to note that while a classical algorithm finds all coefficients $x_i$ for $x$, the HHL algorithm finds the quantum state $\ket{x}=\sum_{i=1}^{N}{x_i\ket{i}}$. Obtaining values for all $x_i$ takes $O(N)$ repetitions; this observation nullifies the speed-up the quantum algorithm has over the classical counterparts. Hence, the HHL algorithm is most useful when used as a subroutine carrying out an intermediate step in a larger process where the quantum state $\ket{x}$ is consumed by the next subroutine in the process.

\subsubsection{Quantum Random Access Memory}
Most quantum algorithms assume parallel access to amplitude encoded states; this is performed using a quantum random access memory or QRAM \cite{giovannetti2008quantum}. The classical RAM takes a memory address as input and returns the data stored at that address. A QRAM performs a similar operation using qubit registers. The input register contains a superposition of addresses $\sum_{a}c_a\ket{a}$ and the output register contains a superposition of the data at those addresses $\sum_{a}c_a\ket{a}\ket{D_a}$. While a classical RAM queries $N$ addresses in $O(N)$, QRAM performs the operation in $O(\log N)$.

\subsection{Quantum Annealing}
Quantum annealing is a metaheuristic optimization algorithm that leverages quantum effects to solve quadratic unconstrained binary optimization (QUBO) problems that deal with optimizing functions of the form:
\begin{align}
C(x_1,x_2,...,x_n) = \sum_{i}a_ix_i + \sum_{i,j}b_{i,j}x_ix_j
\end{align}
where $a_i \in \mathbb{R}$, $b_{i,j} \in \mathbb{R}$, $x_i \in \{0,1\}$. A wide range of problems can be mapped to QUBO and then solved by quantum annealers which are special-purpose quantum computers specifically built to perform quantum annealing.

The Ising model is used in physics to represent a large variety of systems. It was originally proposed to model magnetic materials where every molecule has a spin that can align or anti-align with an applied magnetic field \cite{bian2010ising}. The Hamiltonian of the system representing its energy is given by:
\begin{align}
H = \sum_{i}h_is_i + \sum_{i,j}J_{i,j}s_is_j
\end{align}
where $s_i \in \{-1, +1\}$ is the spin of the $i^{th}$ molecule, $h_i$ is the strength of the magnetic field at the $i^{th}$ molecule, and $J_{i,j}$ is the strength of the interaction between neighboring \mbox{spins $i$ and $j$}. From equations (24) and (25), it can be seen that the QUBO problem convenient maps onto the Ising Hamiltonian with the mapping $s_i=2x_i-1$.

Quantum annealing works as follows. An initial Hamiltonian $H_0$ that is easy to construct is chosen. The system is evolved under a time-dependent Hamiltonian given by:
\begin{align}
H(t) = (1-t)H_0 + tH_f
\end{align}
where $t$ is gradually changed from $0$ to $1$ and the final Hamiltonian $H_f$ is the same as in equation (25). At $t=0$, the system starts in the ground state of $H_0$. According to the quantum adiabatic theorem, the system remains in the ground state of the instantaneous Hamiltonian $H(t)$ throughout its evolution provided it is changed sufficiently slowly \cite{ambainis2004elementary}. At $t=1$, the final Hamiltonian of the system will encode the solution to the problem.

Quantum annealing should not be conflated with the more general adiabatic quantum computing. Quantum annealing specifically solves optimization problems; adiabatic quantum computing is a model of quantum computing that is equivalent to a universal quantum computer. The Hamiltonians used in quantum annealing are classical Hamiltonians, while adiabatic quantum computing uses quantum Hamiltonians that have no classical counterparts \cite{biswas2017nasa}. For a more comprehensive treatment of adiabatic quantum computing, we refer the reader to \cite{albash2018adiabatic}.

\subsection{Quantum Algorithms for Machine Learning}
In this section, we explore how the background and subroutines described in the previous sections can be applied to solve machine learning problems. The common supervised machine learning setting is as follows. The training set consists of $M$ instances $\{X_1,...,X_m,...,X_M\}$ with corresponding labels $\{Y_1,...,Y_m,...,Y_M\}$.  Each $X_m$ is represented by an N-dimensional feature vector $\{x_1^{(m)},x_2^{(m)},...,x_N^{(m)}\}$. Each label $Y_m$ can be either a real value (for regression problems) or a discrete class label (for classification problems). In the \textit{training} phase, the algorithm extracts patterns from the dataset and \textit{learns} a model. In the \textit{inference} phase, this model is used to process unseen instances and predict their corresponding labels.

\subsubsection{k-Nearest Neighbors}
The k-nearest neighbors (KNN) is one of the simplest supervised learning algorithms. To predict the label of a new unseen instance $x_{test}$, the algorithm looks at $k$ instances in the training set that are closest to $x_{test}$ and chooses the class that appears most often in the labels of these \textit{k-nearest neighbors} as the predicted label (for regression, the algorithm assigns the mean value of the k-nearest neighbors as the label). An advantage of KNN is that, unlike many other supervised algorithms, it is non-parametric and makes no assumptions about the data distribution. However, since all computation is deferred, inference can become prohibitively expensive for large training sets. During inference, the distance of the test instance from all other training instances is calculated; this is the most computationally intensive step in the process. Hence, quantum versions of KNN focus on faster evaluation of the distance between two instances.

A{\"\i}meur et al. \cite{aimeur2006machine} propose using the overlap $|\braket{a|b}|$ as computed by the swap test (section 3.1.3) as a measure of similarity between $\ket{a}$ and $\ket{b}$. \mbox{Llyod et al. \cite{lloyd2013quantum}} develop a technique based on the swap test to compute the distance between $\overrightarrow{a}$ and $\overrightarrow{b}$. A state \mbox{$\ket{\psi}=\frac{1}{\sqrt{2}}(\ket{0,a}+\ket{1,b})$} is constructed by setting up an ancilla. A second state given by  \mbox{$\ket{\phi}=\frac{1}{\sqrt{Z}}(|\overrightarrow{a}|\ket{0}-|\overrightarrow{b}|\ket{1})$} is constructed where $Z=|\overrightarrow{a}|^2+|\overrightarrow{b}|^2$. Using $\braket{x|x}=|\overrightarrow{x}|^{-1}|\overrightarrow{x}|$, the authors make the observation that $|\overrightarrow{a}-\overrightarrow{b}|^2=Z|\braket{\phi|\psi}|^2$. With this, the distance between $\overrightarrow{a}$ and $\overrightarrow{b}$ can be retrieved using a swap test \cite{schuld2015introduction}. The authors use the above technique to implement the nearest-centroid classification algorithm in which the centroids of all training instances belonging to each class are precomputed; during inference, a class label is assigned to the test instance by calculating its distance from all centroids. Given a training set of $M$ instances, this procedure solves the problem of classifying an $N$-dimensional test vector into one of several classes in $O(log(MN))$ compared to $O(MN)$ required by classical algorithms.

Wiebe et al. \cite{wiebe2014quantum} argue that the nearest-centroid classification presented above can perform poorly in practice since training instances are often embedded in complicated manifolds, and the centroids may lie outside these manifolds. They propose two fast methods for computing the distance between vectors based on an alternative representation of classical information in quantum states, amplitude amplification, and Durr-Hoyer minimum finding \cite{durr1996quantum}.

\subsubsection{Support Vector Machines}
Support vector machines (SVM) \cite{cortes1995support} is a popular classification algorithm that determines the optimal hyperplane that separates instances of two classes in the training data and classifies test instances based on which side of the separating hyperplane they lie on. Given training instances $\{(x_1,y_1),...,(x_i,y_i),...,(x_M,y_M)\}$ where $x_i \in \mathbb{R}^N$ and $y_i \in \{-1, +1\}$, the algorithm learns an $N$-dimensional hyperplane given by $w = [\beta_1,\beta_2,...,\beta_N]^T$ that separates the instances of the two classes with maximum margin. Classification of a test instance $x_t$ is performed as:
\begin{align}
y_t = sign(w^Tx_t + b)
\end{align}
Since the instances may not be strictly separable, slack variables are introduced that provide a \textit{soft} margin by allowing some data points to violate the margin criterion. This formulation is known as the maximum margin classifier or support vector classifier; this however still requires the data points to be \textit{linearly} separable. SVMs overcome this limitation of linear separability by what is known as the \textit{kernel trick} which transforms the feature space into a new, higher-dimensional feature space. Instances that were not linearly separable in the original feature space may be linearly separable in the new feature space. Mathematically, the kernel trick generalizes the dot product between two feature vectors $\braket{x_i,x_j}$ by a kernel function $k(x_i,x_j)$. Different kernels can be chosen depending on the data distribution. Training an SVM involves solving the quadratic programming problem \cite{press2007numerical}:
\begin{align}
&minimize \quad E=\frac{1}{2}\sum_{nm}\alpha_n\alpha_mt_nt_mk(x_n,x_m)-\sum_n\alpha_n, \\ \nonumber
&subject \, to \quad \, 0 \leq \alpha_n \leq C, \\ \nonumber
&and \qquad \quad \; \, \sum_n\alpha_nt_n=0
\end{align}
where $\alpha_n \in \mathbb{R}$, $C$ is the regularization parameter, and $k(\cdot,\cdot)$ is the kernel function.

Anguita et al. \cite{anguita2003quantum} observe that training support vector machines may be a hard problem and propose a quantum variant that uses Durr and Hoyer's minimum finding \cite{durr1996quantum} based on Grover's algorithm to solve the optimization problem. Rebentrost et al. \cite{rebentrost2012quantum} suggest computing inner products using a quantum method based on an approach similar to the one discussed in section 3.3.1 which leads to an exponential speed-up with respect to the dimension of the feature vector $N$. They also describe a least-squares reformulation of the SVM algorithm with slack variables $e_j$ that converts the quadratic optimization problem into a problem of solving a system of linear equations which leads to an additional exponential speed-up in terms of the number of training instances $M$:
\begin{align}
y_j(\overrightarrow{w} \cdot \overrightarrow{x_j}+b) \geq 1 \rightarrow (\overrightarrow{w} \cdot \overrightarrow{x_j}+b)=y_j-y_je_j
\end{align}

As shown in section 3.2, quantum annealing is particularly well-suited for solving optimization problems. Willsch et al. \cite{willsch2020support} demonstrate a practical implementation of training SVMs on the commercially available D-Wave DW2000Q quantum annealer by formulating it as a QUBO problem.

\subsubsection{Neural Networks}
Artificial neural networks or simply neural networks were originally inspired from biological neural networks that model the activity of neurons in human brains. The basic building block of a neural network is the neuron, also called node or perceptron, that maps the input $x \in \mathbb{R}^N$ to the output $y \in \mathbb{R}$ as follows\footnote{This is the general form used in modern feedforward neural networks. The original perceptron used a step activation function and produced only binary outputs 0 and 1.}:
\begin{align}
y = g(\sum_{i=1}^{N}{w_ix_i} + b)
\end{align}
where $w_i \in \mathbb{R}$, $b \in \mathbb{R}$, and $g: \mathbb{R} \rightarrow \mathbb{R}$ is the activation function. The outputs of some neurons can be fed as inputs to other neurons thus creating layers within the network.

Even though neural networks were amongst the first machine learning algorithms to be proposed, their research stagnated for several decades from 1940s to 1980s due to the inherent difficulty and large computational power required to train them. They returned to popularity after the introduction of backpropagation \cite{rumelhart1986learning} which eased these problems by offering a faster method for training. In the last decade, with GPUs and cloud computing providing cheaper access to massive computational power, neural networks have dwarfed other learning algorithms\footnote{Many informal texts now relegate all other learning algorithms to the category \textit{conventional machine learning}.} to become one of the biggest success stories of modern computers finding applications in various industries including healthcare, manufacturing, finance, analytics etc. to solve problems in image processing, computer vision, natural language processing, predictive modeling, and many other areas. However, even today, significant resources are required to train neural networks, and training times for research and industrial problems can run into weeks or even months. 

An obvious difficulty arises in considering quantum computation as a means for implementing neural networks. Quantum computation (and indeed quantum mechanics itself) is a theory fundamentally based on linear transformations, while an important practical advantage neural networks enjoy over many other learning algorithms is that they can model non-linear data distributions. Bringing non-linearity into quantum algorithms is a non-trivial task \cite{cao2017quantum}. However, classical neural networks do make heavy use of linear algebra, and the inherent randomness of quantum mechanical effects can be leveraged to automatically introduce noise in the training process to improve model robustness - something that needs to be done purposefully in classical training \cite{allcock2018quantum}. Numerous neural network architectures have emerged in recent times to tackle problems belonging to a wide range of supervised, unsupervised, and reinforcement learning tasks \cite{goodfellow2016deep}. Research in this field has been scattered with different proposals addressing narrow problems in piecemeal style. We present here a select subset of these proposals.

Most early work on \textit{quantizing}\footnote{Developing a quantum alternative to a classical computation technique is often referred to as \textit{quantizing} it although we prefer this term is used sparingly.} neural networks focussed on Hopfield networks \cite{hopfield1982neural} which differ from the neural networks presently used in practice. In a Hopfield network, all neurons have undirected connections with all other neurons as opposed to feed-forward networks that are organized as layers; also, each neuron outputs a binary 0 or 1 instead of a real number. Hopfield networks are used to model associative memories which allow retrieval of data based on content rather than addresses. Kak \cite{kak1995quantum} introduces the idea of \textit{quantum neural computation} and Perus \cite{peruvs2000neural} describes a \textit{quantum associative network} by drawing analogies between Hopfield networks and quantum information processing. Behrman et al. \cite{behrman2000simulations} present a quantum realization of neural networks by showing that a single quantum dot\footnote{Quantum dots are nanometre-scale semiconductor particles.} molecule can act as a recurrent quantum neural network. More recently, Rebentrost et al. \cite{rebentrost2018quantum} present a technique based on quantum Hebbian learning and fast quantum matrix inversion to train Hopfield networks.

Boltzmann machines \cite{ackley1985learning}, closely related to Hopfield networks, are stochastic generative networks that can learn a probability distribution over a set of inputs. They are trained by adjusting the interconnection weights between the neurons so that the thermal statistics of the system as described by the Boltzmann-Gibbs distribution reproduces the statistics of the data \cite{biamonte2017quantum}. Boltzmann machines can be conveniently represented by an Ising model whose spins encode features and interactions encode statistical dependencies between the features. In a restricted Boltzmann machine, connections exist only between neurons belonging to different layers; this makes them easier to train than fully-connected Boltzmann machines. Restricted Boltzmann machines can be stacked together to form \textit{deep belief networks} that can be used to learn internal representations or can be trained under supervision to perform classification. Training Boltzmann machines is exponentially hard and is performed using approximation techniques like contrastive divergence \cite{hinton2002training}\cite{salakhutdinov2009deep} that rely on Gibbs sampling. Wiebe et al. \cite{wiebe2014quantum} propose quantum methods to efficiently train full Boltzmann machines by preparing a coherent analog of the Gibbs state from which samples can be drawn. Adachi et al. \cite{adachi2015application} investigate an alternative approach of performing the sampling on a D-Wave quantum annealer instead of classical Gibbs sampling.

A different line of research involves developing quantum analogs for classical perceptrons. Schuld et al. \cite{schuld2015simulating} introduce a quantum perceptron model with a step activation function that can be used to develop superposition-based learning schemes in which a superposition of training vectors can be processed in parallel. Kapoor et al. \cite{kapoor2016quantum} develop two quantum techniques for modeling perceptrons; the first provides quadratic speed-up with respect to the number of training instances, and the second provides a quadratic reduction in the scaling of the training time with the margin between the two classes.

Feedforward networks are one of the simplest neural network architectures in which the connections between neurons do not form any loops or cycles. They are usually trained using backpropagation, and the optimization is performed using some variant of gradient descent. Most machine learning architectures used in practice today are based on feedforward networks or their derivatives such as convolutional neural networks or recurrent neural networks \cite{goodfellow2016deep}. Allcock et al. \cite{allcock2018quantum} define an efficient quantum subroutine for robust inner product estimation using QRAM \cite{giovannetti2008quantum} and use it to demonstrate quadratic speed-ups in the size of the network over classical counterparts; they additionally claim that the proposed quantum method naturally imitates regularization techniques like drop-out leading to more robust networks. Farhi et al. \cite{farhi2000quantum} present a general framework for binary classification specifically designed for near-term quantum processors in which the input strings are mapped to computational basis states, and the predicted label is given by the outcome of a Pauli operator measured on a readout qubit. This framework extends to a full quantum neural network that can classify both classical and quantum data. Convolutional neural networks (CNNs) \cite{lecun1998} have achieved great success \cite{krizhevsky2012imagenet} in image classification tasks in recent times. They, however, suffer from the fact that the operation of convolution is computationally expensive. Kerenidis et al. \cite{kerenidis2019quantum} design a quantum CNN based on quantum computation of the convolution product between two tensors. They also propose a quantum tomography\footnote{Quantum tomography is the process by which a quantum state is reconstructed using measurements on an ensemble of identical quantum states.} sampling approach to recover classical information from the network output and a quantum backpropagation algorithm for efficient training of the quantum CNN.

It is important to distinguish between \textit{quantum-enhanced} machine learning that focuses on techniques to implement learning on classical data using quantum computers from \textit{quantum-generalisation} of machine learning algorithms that deals with developing fully quantum algorithms that work with quantum data. This is especially true for neural networks for which active research is underway on both fronts. We restrict this paper to quantum-enhanced techniques and refer the reader to \cite{cao2017quantum}\cite{beer2020training}\cite{amin2018quantum}\cite{wan2017quantum}\cite{Cong_2019} for work on quantum generalisations. The principles of quantum computing have inspired development of new classical randomized algorithms that show exponential speed-ups over conventional algorithms \cite{tang2019quantum}. With these \textit{quantum-inspired} algorithms, the gap between certain classical and quantum algorithms is no longer exponential but polynomial. Arrazola et al. \cite{arrazola2019quantum} provide a study of these algorithms and observe that they work well only under stringent conditions which occur rarely in practice. We do not cover these in this paper.

\section{Discussion}
Quantum computation has made great strides in the last two decades in both theory and practice. A significant corpus of research has emerged in applying the principles of quantum computation to problems across many fields of science and engineering. At the same time, several approaches for physical realization of quantum computers based on superconducting quantum bits, trapped ions, optical lattices, photonic computing, nuclear magnetic resonance etc. have shown promise. However, fundamental challenges remain unresolved on both fronts. While developing quantum algorithms, we must consider the input problem and the output problem \cite{biamonte2017quantum}. The input problem refers to the fact that the cost of reading classical data and encoding it in quantum states can sometimes dominate performance and render the further downstream speed-up irrelevant. The output problem refers to the reverse process of decoding the full classical solution from quantum states. Some important hardware challenges to constructing, operating, and maintaining large-scale quantum computers include achieving longer coherence, greater circuit depth, higher qubit quality, and higher control over qubits. Quantum error correction plays an important role, and it will likely span across hardware and software in the future.

Owing to its roots in quantum physics, research in quantum computing has so far been confined within the purview of the physics community. Although realization of quantum computers in the form of hardware will remain a problem for physicists, we believe this need not be the case when it comes to applying quantum computing to solve machine learning problems. Classical computing and machine learning, like physics and many other fields, serve as prime examples of disciplines where theoretical results were obtained far before technological progress made possible their experimental realizations. Small-scale quantum computers with less than 100 qubits and quantum annealers with around 2000 qubits have been developed and are already being sold commercially \cite{castelvecchi2017ibm}\cite{johnston2013d}. We hope this article serves its purpose as introductory material for interested machine learning researchers and practitioners from various disciplines.

\bibliographystyle{ieeetr}
\bibliography{bibliography}
\end{document}